\documentclass[twocolumn,showpacs,preprintnumbers,amsmath,amssymb,nofootinbib]{revtex4-1}
\usepackage{here}
\usepackage{comment,braket}
\usepackage{epsf}
\usepackage{amsmath}
\usepackage{graphics}
\usepackage{amsfonts}
\usepackage{amssymb}
\usepackage{latexsym}
\usepackage{color}
\usepackage{ulem}
\input{colordvi.tex}
\usepackage{feynmp}

\usepackage{natbib}
\usepackage[pdftex]{graphicx}
\usepackage{bm}
\usepackage[pdftex]{hyperref}
\usepackage[svgnames]{xcolor}
\definecolor{phthaloblue}{rgb}{0.0, 0.06, 0.54}
\hypersetup{
    colorlinks=true,
    linkcolor=phthaloblue,
    citecolor=blue,
    filecolor=blue,
    urlcolor=phthaloblue,
    }

\newcommand{\beq}{\begin{eqnarray}} 
\newcommand{\eeq}{\end{eqnarray}}

\def\({\left(}
\def\){\right)}
\def\[{\left[}
\def\]{\right]}

\def\Mpl{m_{\rm pl}}
\def\lmk{\left(}
\def\rmk{\right)}

\newcommand{\eq}[1]{Eq.~(\ref{#1})}
\newcommand{\bel}[1] {\begin{equation}\label{#1}}
\newcommand{\beal}[1] {\begin{eqnarray}\label{#1}}
\newcommand{\be}{\begin{equation}}
\newcommand{\ee}{\end{equation}}
\newcommand{\bea}{\begin{array}} 
\newcommand{\eea}{\end{array}}
\newcommand{\abs}[1]{\left\vert#1\right\vert}

\newcommand{\vev}[1]{ \left\langle {#1} \right\rangle }

\newcommand{\GeV}{\  {\rm GeV} }
\newcommand{\TeV}{\  {\rm TeV} }

\begin{document}

\title{Cosmic perturbations, baryon asymmetry and dark matter\\
from the minimal supersymmetric standard model
}

\author{Keisuke Harigaya}
\affiliation{School of Natural Sciences, Institute for Advanced Study, Princeton, NJ 08540, USA}

\author{Masaki Yamada}
\affiliation{Institute of Cosmology, Department of Physics and Astronomy, 
Tufts University, Medford, MA  02155, USA}

\date{\today}

\begin{abstract}
Scalar fields in the minimal supersymmetric standard model may have large field values during inflation. Because of approximate global symmetry, it is plausible that the phase directions of them are nearly massless during inflation and obtain quantum fluctuations, which may be the origin of the cosmic perturbations. If perturbations are produced through Q-ball formation, baryon asymmetry and dark matter can be consistently generated. Significant baryon and dark matter isocurvature perturbations are produced, but they are predicted to nearly compensate each other. The lepton asymmetry is much larger than the baryon asymmetry. The scenario predicts local non-Gaussianity of $f_{\rm NL} = 5/3$. Implication to the mass spectrum of supersymmetric particles is discussed. 
\end{abstract}

\maketitle

{\it Introduction.}--
The universe starts from small fluctuations which have grown and collapsed to form galaxies by the gravitational force. 
The origin of the fluctuations is an elementary question in cosmology and particle physics. 
The observations of the fluctuations of the large scale structure and the cosmic microwave wave background (CMB) have revealed that the fluctuations are nearly Gaussian and scale-invariant~\cite{Bennett:2012zja,Aghanim:2018eyx}, which are most naturally explained by quantum fluctuations of a nearly massless scalar field generated during inflation~\cite{Mukhanov:1981xt,Hawking:1982cz,Starobinsky:1982ee,Guth:1982ec,Bardeen:1983qw}. Almost nothing is known about the scalar field. It may be the inflaton itself or  another scalar field (e.g., a curvaton~\cite{Enqvist:2001zp, Lyth:2001nq, Moroi:2001ct}), may be as heavy as $10^{13}$~GeV or in principle as light as $10^{-24}$~GeV. 

In this Letter, we investigate the possibility that scalar fields in the minimal supersymmetric standard model (MSSM) are responsible for the fluctuations. 
The MSSM is one of the best-motivated extensions of the standard model. It achieves successful precise gauge coupling unification~\cite{Dimopoulos:1981yj,Dimopoulos:1981zb,Sakai:1981gr,Ibanez:1981yh,Einhorn:1981sx,Marciano:1981un}, provides a dark matter candidate as the lightest supersymmetric particle (LSP)~\cite{Witten:1981nf,Pagels:1981ke,Goldberg:1983nd}, and can explain the electroweak scale much smaller than the unification scale $\sim 10^{16}$ GeV~\cite{Maiani:1979cx,Veltman:1980mj,Witten:1981nf,Kaul:1981wp}.
In the MSSM, there exist many combinations of scalar fields whose potentials vanish at the renormalizable level and supersymmetric limit, which are called flat directions. 
It is natural to ask if the cosmic perturbations originate from quantum fluctuations of one of those flat directions~\cite{Enqvist:2002rf,Enqvist:2003mr,Kasuya:2003va, Hamaguchi:2003dc, McDonald:2003jk, Riotto:2008gs}, and seek experimental signatures of such a scenario.

With supersymmetry broken during inflation, the flat directions in general obtain soft mass squared terms called Hubble induced masses. Suppose
a flat direction $\phi$ obtains a nagative Hubble-induced mass squared and hence a large field value during inflation~\cite{Dine:1995uk, Dine:1995kz}.
The flat direction is stabilized by a higher dimensional term in the super-potential $W = \lambda \phi^{n}$. The potential of $\phi$ is 
\begin{align}
V \sim - H^2 |\phi|^2 + \lambda^2 |\phi|^{2n-2},
\end{align}
so that the field value of $\phi$ during inflation is $\phi_{\rm inf}\sim (H_{\rm inf}/\lambda)^{1/(n-2)}$, where $H_{\rm inf}$ is the Hubble scale during inflation. The radial direction has a mass as large as the Hubble scale and its quantum fluctuations are suppressed. A special case of nearly vanishing Hubble induced masses is discussed in~\cite{Enqvist:2002rf,Enqvist:2003mr,Kasuya:2003va,Hamaguchi:2003dc}. 
Up to this point the scalar potential of $\phi$ possesses an $U(1)$ symmetry (which is an R symmetry).
It is conceivable that even if other possible higher dimensional operators are included, the $U(1)$ symmetry is approximately maintained, like the baryon and lepton symmetry. 
The phase direction of $\phi$, which we denote as $\theta$, is then almost massless and obtains quantum fluctuations $\delta \theta \sim H_{\rm inf} / (2 \pi \phi_{\rm inf})$ during inflation. We investigate if the fluctuations of the phase direction alone can explain the cosmic perturbations.
This scenario is examined in~\cite{McDonald:2003jk} neglecting important dissipation effects explained below, while \cite{Riotto:2008gs} predicts too large non-Gaussianity.

After inflation, the inflaton, which is different from $\phi$, begins oscillation. We assume that the sign of the Hubble induced mass squared of $\phi$ remains negative. The field value of $\phi$ tracks the minimum, $\phi \sim (H/\lambda)^{1/(n-2)} $~\cite{Dine:1995uk, Dine:1995kz,Harigaya:2015hha}. When the Hubble scale drops below the soft mass of $\phi$ given by the supersymmetry breaking at the vacuum, $m_\phi$, the flat direction begins oscillation around the origin. 
At this point, the irreducible explicit $U(1)$ symmetry breaking from the gravitino mass $m_{3/2}$,
\begin{align}
\Delta V \sim m_{3/2} \lambda \phi^n + {\rm h.c.},
\end{align}
which gives a non-zero potential energy of $\theta$,
is no more negligible.
The fluctuation of $\theta$, together with its potential, gives rise to the fluctuation of the energy density and hence may sources the cosmic perturbations~\cite{McDonald:2003jk,Riotto:2008gs}. 
This seems challenging for a MSSM flat direction, since it has $O(1)$ gauge couplings. As is shown in~\cite{Mukaida:2014yia}, such a field is rapidly dissipated into thermal plasma and never dominates the universe. If the curvature perturbation of the universe originates from fluctuations of a subdominant component, the fluctuations must be large and  the resultant perturbation is highly non-Gaussian~\cite{Lyth:2002my}, which is incompatible with the observations~\cite{Aghanim:2018eyx}.

The rapid dissipation is due to a passage of the field near the origin, which is avoided if $\phi$ circulates by a kick from large enough $\Delta V$.
The circular motion corresponds to global charge asymmetry of $\phi$.
It is temping to identify the charge asymmetry with baryon asymmetry~\cite{Dine:1995uk, Dine:1995kz}, but too large baryon asymmetry is produced if $\phi$ dominates the universe. Moreover, since the energy of the phase direction is comparable to or smaller than that of the radial direction, too large baryon isocurvature perturbation is produced~\cite{Lyth:2002my}.

These problems are evaded if Q-balls form. For some flat directions, the potential of the radial direction is shallower than a quadratic one.
Then as $\phi$ oscillates, it develops instability and non-topological solitons called Q-balls~\cite{Coleman:1985ki} are formed~\cite{Coleman:1985ki, Kusenko:1997si,Enqvist:1997si,Enqvist:1998en,Kasuya:1999wu}, into which the energy of the phase direction is converted. The dissipation rate of Q-balls is limited by the Pauli-blocking near the surface of Q-balls~\cite{Cohen:1986ct}. Q-balls are more long-lived than the radial direction and can dominate the universe.
If the Q-balls have  vanishing baryon charges but non-zero lepton charges, appropriate amount of baryon asymmetry is generated from partial decay of the Q-balls before the electroweak phase transition and the sphaleron effect~\cite{Kuzmin:1985mm}.

In the following we describe the detail of the scenario as well as its cosmological and astrophysical signatures.

{\it Curvature perturbation from Q-balls}--
When $\phi$ begins oscillation, the explicit breaking in $\Delta V$ induces angular motion of $\phi$ and produces asymmetry of the global charge of $\phi$~\cite{Affleck:1984fy},
\begin{align}
n_\phi \sim \epsilon m_\phi \phi_{\rm osc}^2,~~\epsilon \equiv \theta_{\rm osc} \frac{m_{3/2}}{m_\phi}.
\end{align}
The field value of $\theta$ is defined so that $\theta =0 $ is the minimum of the potential $\Delta V$. The subscript ``osc'' represents the value at the beginning of the oscillation. 
Since  $\theta$ fluctuates, the charge asymmetry also fluctuates. 
The asymmetry is approximately conserved afterward since $\Delta V$ becomes negligible as the amplitude of the oscillation of the radial direction decreases due to the cosmic expansion.
Once the Q-balls are formed, most of the asymmetry is stored in them.
As we will see, $\epsilon$ is required not to be much below unity. This is naturally the case if the potential of $\phi$ around the field value $\phi_{\rm osc}$ is dominantly given by gravity mediation,
\begin{align}
V(\phi) = m_{\phi}^2 |\phi|^2\left( 1 + K {\rm ln} \frac{|\phi|^2}{\Mpl^2}\right) ,
\end{align}
where $K$ is negative and $m_\phi \sim m_{3/2}$. The logarithmic term comes from the renormalization running of the soft mass. 
$\abs{K}= O(0.1\mathchar`-0.01)$ if soft masses are of the same order, while can be as small as $10^{-4}$ if gaugino masses are much smaller than scalar masses at a high energy scale. 
The following computation is also applicable to gauge mediation as long as the potential energy is dominated by gravity mediated one around $\phi = \phi_{\rm osc}$.
The parameter $m_{\phi}$ may be below the electroweak scale.

The number density of the Q-balls is determined by the instability scale $\sim \abs{K}^{-1/2} m_\phi^{-1}$, almost independent of $\theta_{\rm osc}$~\cite{Hiramatsu:2010dx}, and
does not fluctuate.  The charge $Q$ and the mass $m_Q$ of individual Q-balls (averaged in each Hubble patch) fluctuate as they depend on $\theta_{\rm osc}$ via $\epsilon$~\cite{Kusenko:1997si, Kasuya:2000wx, Hiramatsu:2010dx, Doddato:2011fz},
\begin{align}
Q \simeq  0.02 \epsilon \lmk \frac{\abs{K}}{0.1} \rmk^{1/2} \lmk \frac{\phi_{\rm osc}}{m_\phi} \rmk^2 \label{Q},~~
m_Q \simeq m_\phi Q, 
\end{align}
where we assume $\epsilon \gtrsim 0.01$.
We comment on the case with smaller $\epsilon$ as well as a subtle issue for $\epsilon \sim 1$ later.
This leads to the fluctuations of the energy density $\rho_Q$ and the decay rate $\Gamma_Q$ of the Q-balls,
\begin{align}
\label{eq:energy_decay}
\rho_Q (a) \simeq  \epsilon m_\phi^2 \phi_{\rm osc}^2 \lmk \frac{a_{\rm osc}}{a} \rmk^3,~
\Gamma_Q \simeq g \frac{R^2 m_\phi^3}{24 \pi Q}  \ \ (\propto \epsilon^{-1} ), 
\end{align}
where $R \sim 7/\sqrt{\abs{K}} m_\phi$ 
is the typical size of the Q-balls and $g$ $(\sim 100)$ is the number of degrees of freedom that couples to the Q-balls~\cite{Cohen:1986ct, Kamada:2012bk}. 
The decay rate of Q-balls is suppressed by the Pauli-blocking effect, guaranteeing the long-lifetime of Q-balls.

At the time of the production, the Q-balls are subdominant component of the universe.
Since the Q-balls are long-lived, they eventually dominate the energy density of the universe and then decay into standard model particles 
at a temperature 
\begin{align}
 T_{\rm dec} &\simeq   \lmk \frac{10}{\pi^2 g_* (T_{\rm dec})} \rmk^{1/4} \sqrt{\Gamma_Q \Mpl} \\
 &\simeq  3.6 \GeV \lmk \frac{m_\phi}{10 \TeV} \rmk^{1/2} \lmk \frac{1.3 \times 10^{24}}Q \rmk^{1/2} \propto \epsilon^{-1/2}, \nonumber
\end{align}
where $g_*$ is the effective number of relativistic particles and we take $\abs{K} =0.01$. 
The decay rate and the energy density of the Q-balls are modulated by the fluctuation of $\epsilon$ (or $\theta_{\rm osc}$), which sources the curvature perturbation of the universe.

We compute the magnitude of the curvature perturbation by the $\delta N$ formalism~\cite{Sasaki:1995aw,Wands:2000dp,Lyth:2004gb} with the following history.
After inflation and reheating complete,
1) the field $\phi$ begins oscillation. Soon after that, the instability occurs and Q-balls form. 
2)  Since the energy density of the Q-balls decreases slower than that of radiation, the Q-balls eventually dominate the universe and 3) decay afterward  at a temperature $T_{\rm dec}$. 4) The universe reaches a reference temperature $T_f$. It is also possible that  $\phi$ begins oscillation before the completion of the reheating, but the property of the produced curvature perturbation remains the same.

We parametrize the energy density of the Q-balls at the beginning of the oscillation as $\rho \epsilon$, and the decay rate of them as $\Gamma / \epsilon$. Numbers of e-foldings between the stages 1)-4) are
\begin{align}
N_{12} \simeq  {\rm ln}  \frac{ m_\phi^2  \Mpl^2}{\rho \epsilon},~
N_{23} \simeq \frac{1}{3} {\rm ln} \frac{\rho \epsilon}{ \Gamma^2 \Mpl^2 / \epsilon^2},~
N_{34} \simeq  \frac{1}{4} {\rm ln} \frac{ \Gamma^2 \Mpl^2 / \epsilon^2}{T_f^4}.
\end{align}
The total number of e-foldings is then
\begin{align}
N_{\rm tot} = - \frac{1}{2} {\rm ln} \epsilon + {\rm const.}
\end{align}
To the first order in $\delta \epsilon$,
the curvature perturbation produced by the Q-balls is
\begin{align}
\zeta = \delta N = - \frac{1}{2} \frac{\delta \epsilon}{ \epsilon}.
\end{align}
The parameter $\epsilon$ is proportional to the field value of $\theta$. The observed curvature perturbation ${\cal P}_\zeta \simeq 3 \times 10^{-10}$~\cite{Akrami:2018odb} requires that
\begin{align}
H_{\rm inf} \simeq 1 \times 10^{13}~{\rm GeV} \frac{\phi_{\rm inf}}{10^{17}~{\rm GeV}}.
\end{align}
The contribution to the curvature perturbation through the modulation of the decay rate of the Q-balls is similar to the modulated reheating scenario~\cite{Dvali:2003em, Kofman:2003nx}. 
There is no spectator field in our scenario. The decay rate is modulated by the property of the Q-balls themselves.

The spectral index is given by~\cite{Sasaki:1995aw,Lyth:1998xn}
\begin{align}
n_s = 1 - 2 \epsilon_{\rm inf} + 2 \eta_\theta,
\end{align}
where $\epsilon_{\rm inf}$ is the first slow-roll parameter of inflation and $\eta_\theta$ ($= m_\theta^2 / H_{\rm inf}^2$) is determined by the mass of the phase direction $m_\theta$. The red-tilted spectrum $n_s = 0.96-0.98$ requires $\epsilon_{\rm inf} = O(10^{-2})$ or $m_{\theta}^2 = - O(10^{-2}) H_{\rm inf}^2$. The former requires large field inflation~\cite{Lyth:1996im}, while the latter requires explicit breaking of the approximate $U(1)$ symmetry. In fact, if $\phi_{\rm inf}$ is close to the cut off scale, we expect that some higher dimensional terms in the Kahler-potential or super-potential are not negligible and give $\theta$ a mass not much below the Hubble scale.

The local non-Gaussianity is parametrized by $f_{\rm NL}$ defined by
\begin{align}
\zeta = \zeta_g +  \frac{3}{5} f_{\rm NL} \zeta_g^2,
\end{align}
where $\zeta_g$ is a Gaussian perturbation.
In our model,
\begin{align}
f_{\rm NL} = \frac{5}{3},
\end{align}
which may be detected by future observations of galaxy distributions~\cite{Alvarez:2014vva,Dore:2014cca} or 21cm lines~\cite{Munoz:2015eqa}.

If $\epsilon \lesssim 0.01$, Q-balls and anti-Q-balls of the almost same charges and energy densities are produced, invalidating \eq{Q}~\cite{Kasuya:2000wx, Multamaki:2002hv, Hiramatsu:2010dx}. 
The fluctuation of $\theta$ only perturbs the charges and number densities of the subdominant component of the Q-balls, producing too large non-Gaussianity. (The threshold value of $\epsilon$ for the formation of anti-Q-balls may depend on $\abs{K}$, but the precise determination of the value requires a detailed numerical simulation of Q-ball formation and is beyond the scope of this Letter.)

We have implicitly assumed that the dynamics of the radial direction is independent of $\epsilon$, which is the case for $\epsilon < 1$. 
We expect that the dynamics is affected for $\epsilon \sim 1$ as the potential energy of the phase direction is comparable to that of the radial direction. For example, the beginning of oscillation  as well as the size of Q-balls may depend on $\epsilon$, giving $O(\epsilon)$ corrections to above formulae. For simplicity we assume $\epsilon$ is not close to unity, and leave the investigation of $\epsilon \sim 1$ for future works.

{\it Baryon asymmetry.}--
If the Q-balls do not have baryon nor lepton charges, or have a vanishing $B-L$ charge and decay before the electroweak phase transition, they do not produce baryon asymmetry. Baryon asymmetry can be produced after the Q-balls decay at a low temperature. Possible scenarios include the electroweak baryogenesis~\cite{Cohen:1990py, Carena:2008vj} and the baryogenesis from neutrino oscillation~\cite{Asaka:2005pn}. No baryon isocurvature perturbation is produced in these cases.

The minimal and more interesting possibility is that the Q-balls produce baryon asymmetry.
If all amount of asymmetries in the Q-balls  are released into baryons or quarks via the decay of the Q-balls, too much baryon asymmetry is produced.
We instead consider Q-balls with vanishing baryon charges but non-zero lepton charges $q_L$ (i.e.~L-balls), which decay after the electroweak phase transition. Then only leptons emitted from the Q-balls before the electroweak phase transition are converted into baryons by the sphaleron process~\cite{Kuzmin:1985mm}. Assuming that the thermal bath is dominated by particles produced by the Q-balls rather than the inflaton produced ones, the amount of the baryon asymmetry $n_b$ normalized by the entropy density $s$ is
\begin{align}
\label{eq:nb}
\frac{n_b}{s} 
&\simeq \left. \frac{28}{79} \abs{q_L} \frac{Q \Gamma_Q n_Q(t) t}{4 \rho_Q(t) / 3 T_{\rm dec}} \right\vert_{T = T_{\rm EW}} 
\nonumber \\
&\simeq \abs{q_L} \frac{140g_{*}(T_{\rm dec}) }{79 g_* (T_{\rm EW})} \frac{T_{\rm dec}^5}{m_{\phi} T_{\rm EW}^4} 
\nonumber \\
&\simeq 9 \times 10^{-11} \abs{q_L} \left( \frac{T_{\rm dec}}{3.6 {\rm GeV}} \right)^5 \frac{10~{\rm TeV}}{m_{\phi}},
\end{align}
where
$T_{\rm EW}$ ($\simeq 174 \GeV$) is the electroweak phase transition temperature. 
The result explains the observed one if $T_{\rm dec}$ is around a few GeV.

The modulated decay temperature $T_{\rm dec} \propto \epsilon^{-1/2}$ produces 
the baryon isocurvature perturbation,
\begin{align}
S_{\rm B} = - \frac{5}{2} \frac{\delta \epsilon}{ \epsilon} = 5 \zeta,
\label{S_B}
\end{align}
which is correlated with the curvature perturbation. This seems to be excluded by the CMB observations.
However, as we will see, the dark matter isocurvature perturbation naturally compensates the baryon isocurvature perturbation.

The compensation relies on the specific dependence of the baryon asymmetry on $T_{\rm dec}$, which is altered if the inflaton-originated particles dominate the thermal bath of standard model particles at the electroweak phase transition (for which we find $S_B = 4 \zeta$.) To avoid it, the Q-balls must dominate the universe early enough, which requires large $\phi_{\rm osc}$. To maintain $T_{\rm dec}\sim$ few GeV, $m_{\phi}$ must be large enough,
\begin{align}
\label{eq:mphibound}
m_\phi \gtrsim  10~{\rm TeV} \left(  \frac{|K|}{0.01} \right)^{5/13} \left(  \frac{100}{g}  \right)^{10/39},
\end{align}
where we use Eqs.~(\ref{eq:energy_decay},\ref{eq:nb}).
Such a large scalar soft mass is consistent with the observed Higgs mass~\cite{Okada:1990vk,Ellis:1990nz,Haber:1990aw}.
The bound is relaxed if the inflaton dominantly decays into particles which decouple from standard model particles.
Still, the universe must be Q-ball dominated when the temperature of the standard model bath produced from the Q-balls is around the electroweak scale, requiring 
\begin{align}
\label{eq:mphibound}
m_\phi \gtrsim  300~{\rm GeV} \left(  \frac{|K|}{0.01} \right)^{5/11} \left(  \frac{100}{g}  \right)^{10/33}.
\end{align}

{\it Dark matter.}--
If dark matter abundance is established after the Q-balls decay, the dark matter isocurvature perturbation is absent. This should be the case if baryon asymmetry is also produced after the Q-balls decay. We instead consider the production of the LSP dark matter by the decay of the Q-balls.
The temperature at which the Q-balls decay is around the GeV scale, which is well after the freeze-out of the LSP, or that of the Next-to-LSP (NLSP) if the gravitino is the LSP. The LSP is hence produced non-thermally.

Suppose the LSP is produced by the decay of the Q-balls, and the annihilation of them afterward is negligible. This is the case if the Bino is the LSP or the gravitino is the LSP and the NLSP quickly decays into the gravitino.
The dark matter abundance is proportional to $T_{\rm dec}$, producing a dark matter isocurvature perturbation  $S_{\rm DM} = - (\delta \epsilon/ \epsilon) /2 = \zeta$. Together with the baryon isocurvature perturbation, this scenario predicts too large a matter isocurvature perturbation and is excluded by the CMB observations.

If the LSP is electroweak charged (i.e.~Wino or Higgsino-like), or is the gravitino but the NLSP does not decay into the gravitino quickly and is standard model gauge charged (any superpartners but a Bino-like one), the annihilation of the (N)LSP just after the production diminishes the dark matter abundance,
\begin{align}
\frac{\rho_{\rm DM}}{s} \simeq &  \left. m_{\rm LSP}\frac{H}{ \vev{\sigma v} s} \right\vert_{T = T_{\rm dec}} \\
\simeq&  0.4~{\rm eV} \frac{3~{\rm GeV}}{T_{\rm dec}} \frac{m_{\rm LSP}}{0.7~{\rm TeV}} \left( \frac{m_{\rm (N)LSP}}{0.7~{\rm TeV}}  \right)^2 \frac{0.01 / m_{\rm (N)LSP}^2}{\vev{\sigma v}}. \nonumber
\end{align}
If the LSP is not the gravitino, the LSP should have a mass below TeV.
In gravity mediation, if the LSP is the gravitino, the NSLP is not much heavier than the gravitino and the NLSP decays during the Big-Bang Nucleosynthesis. Such case is excluded unless the sneutrino is the LSP~\cite{Kawasaki:2008qe}. 
In both cases, the resultant dark matter isocurvature perturbation is
\begin{align}
S_{\rm DM} = \frac{1}{2} \frac{\delta \epsilon}{ \epsilon} = - \zeta,
\label{S_DM}
\end{align}
which is $-1/5$ of $S_B$.

{\it Compensated isocurvature perturbations.}--
As dark matter is nearly five times more abundant than baryons, the matter isocurvature
\begin{align}
S_m = \frac{\Omega_{\rm DM}}{\Omega_m} S_{\rm DM} + \frac{\Omega_{B}}{\Omega_m} S_{B}
\end{align}
nearly vanishes according to Eqs.~(\ref{S_B},\ref{S_DM}). That is, the isocurvature perturbations are compensated with each other and are less-constrained by CMB observations~\cite{Gordon:2002gv}.
In our model the compensation is a natural consequence of the dependence  of the baryon asymmetry and the dark matter abundance on $T_{\rm dec}$, while the compensation in the literatures~\cite{Gordon:2002gv,Harigaya:2014bsa} requires tuning of model parameters.

The upper bound ${\cal P}_{S_m} < 0.001 {\cal P}_\zeta$~\cite{Akrami:2018odb} is satisfied for $4.8 < \Omega_{\rm DM}/\Omega_{B}<5.2$, which is consistent with the measurement of the abundances by the Planck satellite~\cite{Aghanim:2018eyx}  within the $2\sigma$ level.
Future observations can determine the ratio $\Omega_{\rm DM}/\Omega_B$ with an absolute uncertainty of $0.02$~\cite{Baumann:2017gkg}, and probe ${\cal P}_{S_m}/ {\cal P}_\zeta$ as small as $0.0002$~\cite{Finelli:2016cyd}.
The scenario can be tested by the future observations unless $4.9 < \Omega_{\rm DM}/\Omega_{B}<5.1$.

The prediction is altered if there exists another subdominant component of dark matter. A well-motivated example is the QCD axion~\cite{Peccei:1977hh,Peccei:1977ur,Weinberg:1977ma,Wilczek:1977pj} which solves the strong CP problem~\cite{tHooft:1976rip}. If the oscillation of the axion begins at $T> T_{\rm dec}$, the axion dark matter abundance depends on a positive power of $T_{\rm dec}$. This produces a dark matter isocurvature perturbation correlated with the curvature perturbation and reduces the matter isocurvature perturbation. Although the matter isocurvature perturbation is no longer uniquely predicted, one may check the consistency of the scenario once the decay constant and the abundance of the QCD axion are measured.

{\it Lepton assymetry.}--
The lepton asymmetry is large and negative,
\begin{align}
\frac{n_{L_i}}{s} \simeq  - |q_{L_i}|\frac{3T_{\rm dec}}{4m_{\phi}} 
\simeq - 3 \times 10^{-4} |q_{L_i}| \frac{T_{\rm dec}}{3.6 {\rm GeV}} \frac{10~{\rm TeV}}{m_{\phi}},
\end{align}
where $i$ is the generation index and $q_{L_i}$ is the $i$th generation lepton number of the Q-balls.
Assuming that the asymmetry is equally distributed among three generations by neutrino oscillation~\cite{Dolgov:2002ab,Wong:2002fa,Mangano:2011ip}, the upper bound $|n_{L_i}/s| \lesssim 0.01$~\cite{Oldengott:2017tzj} requires $m_\phi \gtrsim100$ GeV.

If $m_{\phi}$ is $O(100)$ GeV, the large negative lepton asymmetry increases the abundance of anti-electron neutrinos and hence the neutron-proton ratio, leading to larger Helium abundance produced by the Big-Bang Nucleosynthesis. Since the recombination is more effective, more baryon asymmetry is required to fit the CMB spectrum, helping the compensation. It will be interesting to check if future observations can probe this scenario. 
Note that $m_{\phi}=O(100)$ GeV violates the bound~(\ref{eq:mphibound}) and requires that the inflaton decays into a hidden sector.

The lepton asymmetry increases the effective number of neutrinos $N_{\rm eff}$~\cite{Lesgourgues:1999wu},
\begin{align}
\Delta N_{\rm eff} = \frac{3698 \pi^2}{105} \sum_i \lmk \frac{n_{L_i}}{s} \rmk^2.
\end{align}
Taking into account the above mentioned upper bound, $\Delta N_{\rm eff}\ll 1$. If a particular combination of $L_i$ is produced, however, $|L_e| \ll |L_{\mu,\tau}|$ and $\Delta N_{\rm eff}$ may be large~\cite{Barenboim:2016shh}. For $T_{\rm dec} / m_\phi = 0.1$, $\Delta N_{\rm eff} \simeq 0.5$, which is favored to ameliorate so-called the $H_0$ tension~\cite{Mortsell:2018mfj, Riess:2019cxk}.

{\it Summary and Discussion.}--
We have investigated the possibility that the curvature perturbation of the universe arises from the fluctuation of the phase direction of a flat direction in the MSSM.
The curvature perturbation is given by the fluctuation of the energy density and the decay rate of Q-balls formed from the flat direction.

Baryon asymmetry can be produced by the decay of the Q-balls.
The LSP abundance must be determined by the efficient annihilation of supersymmetric particles just after they are emitted from the Q-balls. Assuming that the gravitino is not the LSP, the LSP must have a mass below TeV and Wino- or Higgsino-like. Signals from the annihilation of the LSP in galaxies are expected; see~\cite{Bhattacherjee:2014dya,Rinchiuso:2018ajn} for recent discussions on future prospects.

Our scenario has the following beyond-$\Lambda$CDM parameters; 1) local non-Gaussianity $f_{\rm NL}= 5/3$ and 2) baryon and dark matter isocurvature perturbations nearly compensating with each other. If the inflaton dominantly decays into a hidden sector, 3) lepton asymmetry and 4) effective number of neutrinos can be sizable.

Our analytical estimations assume the parameter $\epsilon$ not close to 1. Corrections as $\epsilon$ approaches 1 can change the predictions by $O(1)$ factors. Full investigation of the scenario will require numerical computations.

We assume that only L-balls are formed.
One may naively expect that in $SU(5)$ unified theories, when a flat direction to form L-balls has a large field value, $SU(5)$ partners of the flat direction also have large field values to form B-balls. This is not necessarily the case. For example, we may consider the $\bar{d}_1 L_1 Q_2$ flat direction, where the subscripts are generation indices. The $SU(5)$ partners $L_1L_1\bar{e}_2$ and $\bar{d}_1\bar{d}_1 \bar{u}_2$ identically vanish. We may also use the $L H_u$ direction if the lightest neutrino is nearly massless. The $SU(5)$ partner of the direction involves a heavy colored Higgs and cannot be excited. Other possibilities include embedding of quarks and leptons to different multiplets of $SU(5)$, or orbifold GUTs~\cite{Kawamura:2000ev,Hall:2001pg}.

The scenario gives several interesting predictions. Confirmation of the predictions will suggest that supersymmetry plays more significant roles than expected:
In addition to precise gauge coupling unification and the small electroweak scale,
supersymmetry may further provide the seed of the large scale structure, baryon asymmetry and dark matter, which are essential ingredients of the origin of the present universe.

\section*{Acknowledgement}
KH thanks Fuminobu Takahashi and Benjamin Wallisch  for useful discussion.
This work was supported in part by the Director, Office of Science, Office of High Energy and Nuclear Physics, of the US Department of Energy under Contracts DE-SC0009988 (KH).

\bibliography{reference}

\end{document}